\documentclass[journal]{IEEEtran}
\usepackage{graphicx}
\usepackage{subfigure}
\usepackage{algorithm}
\usepackage{algorithmic}
\usepackage{cite}
\usepackage{amsfonts}
\usepackage{psfrag}
\usepackage{footnote}
\usepackage{amsmath}
\usepackage{multirow}
\usepackage[switch,pagewise]{lineno}

\newtheorem{lemma}{\textbf{Lemma}}
\newtheorem{theorem}{\textbf{Theorem}}
\newtheorem{corollary}{\textbf{Corollary}}
\newtheorem{definition}{\textbf{Definition}}

\begin{document}

%\linenumbers
\title{Stability and Delay Analysis of EPON Registration Protocol}

\author{Qingpei Cui,
        Tong Ye,~\IEEEmembership{Member,~IEEE,}
        Tony T. Lee,~\IEEEmembership{Fellow,~IEEE,}
        Wei Guo,~\IEEEmembership{Member,~IEEE}
        and Weisheng Hu,~\IEEEmembership{Member,~IEEE}
\thanks{This work was supported by the National Science Foundation of China under Grants 61271215, 61001074, 61172065, 61071080, and 60825103, in part by Qualcomm Corporation Foundation, in part by the 973 program (2010CB328205, 2010CB328204), and in part by Shanghai 09XD1402200.}
\thanks{The authors are with the State Key Laboratory of Advanced Optical Communication Systems and Networks, Shanghai Jiao Tong University, Shanghai 200030, China (e-mail: \{cuiqingpei, yetong, ttlee, wguo, wshu\}@sjtu.edu.cn.
}% <-this % stops a space
}

\markboth{IEEE Transactions on Communications,~Vol.~x, No.~x, October~2013}%
{Shell \MakeLowercase{\textit{et al.}}: Bare Demo of IEEEtran.cls for Journals}

\maketitle

\begin{abstract}
The Ethernet passive optical network (EPON) has recently emerged as the mainstream of broadband access networks. The registration process of EPON, defined by the IEEE 802.3av standard, is a multi-point control protocol (MPCP) within the media access control (MAC) layer. As with other contention-based channel access methods, such as ALOHA and CSMA, stability and delay are critical issues concerning the performances of implementing the protocol on systems with finite channel capacity. In this paper, the registration process of an EPON subscriber, called optical network units (ONUs), is modeled as a discrete-time Markov chain, from which we derive the fundamental throughput equation of EPON that characterizes the registration processes. The solutions of this characteristic equation depend on the maximum waiting time. The aim of our stability analysis is to pinpoint the region of the maximum waiting time that can guarantee a stable registration throughput and a bounded registration delay. For a maximum waiting time selected from the stable region, we obtain the expression of registration delay experienced by an ONU attempting to register. All analytic results presented in this paper were verified by simulations.
\end{abstract}


\IEEEpeerreviewmaketitle

\section{Introduction}\label{introduction}
\IEEEPARstart{D}{espite} the growth of Internet traffic and the telecommunications backbone at an unprecedented pace, the access network between end-users and the core network remains the bottleneck for broadband integrated services. The Ethernet passive optical network (EPON) has recently emerged as one of the most promising candidates for broadband access networks. Combining Ethernet technologies and optical fiber infrastructures, the EPON is simple, cost-effective, and capable of supporting various kinds of bandwidth-intensive services, such as video-on-demand (VoD), distance learning, and video conferencing\cite{1kramer2005}. Currently, this widespread technology for fiber-to-the-home (FTTH) applications has become the mainstream of the broadband access market \cite{2kramer2002,3wang2012china,4shinohara2005broadband,5tanaka2010ieee}.

\begin{figure}[t]
\centering
\includegraphics[width=0.40\textwidth]{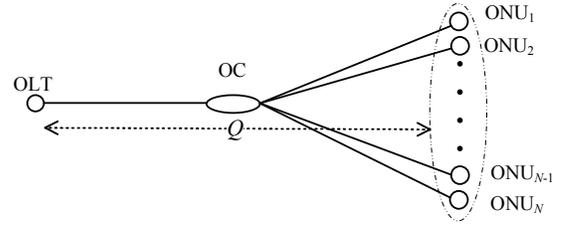}
\caption{Illustration of the EPON network.}\label{fig1}
\end{figure}

The EPON is a point-to-multipoint network, as illustrated in Fig. \ref{fig1}, which consists of an optical line terminal (OLT) at the central office, $N$ optical network units (ONU) at the subscriber side, and a $1:N$ optical coupler (OC) in between. The population of ONUs $N$ is currently 32 or 64, and will extend to 256 or even 512 to meet the ever-increasing demands of FTTH applications \cite{6tran2006low,7chan2010remote,8hajduczenia2010ieee}. Based on an online scheduling of the OLT, the ONUs share the bandwidth of the channel between the OLT and the OC in a time-division-multiplexing (TDM) manner. Thus, the OLT should have information about all active ONUs in advance.

The registration process is a protocol defined by the IEEE 802.3av standard \cite{20standard} that permits the OLT to collect information about those offline ONUs that request access to the network. The OLT initiates a discovery window for each registration process. In the discovery window, all registered online ONUs stop their upstream transmissions, while unregistered ONUs can randomly send the registration requests to the OLT if they want to be connected. As a result, the registration request of an ONU could be ruined if it collides with other randomly generated requests in the discovery window. The failed ONUs tries again by sending registration requests in the next discovery window \cite{1kramer2005,10Bjelica}. The collisions among registration requests reduce the success probability or the throughput of the registration process \cite{1kramer2005,9cui2012throughput}.

\begin{figure*}[t]
\centering
\includegraphics[width=0.80\textwidth]{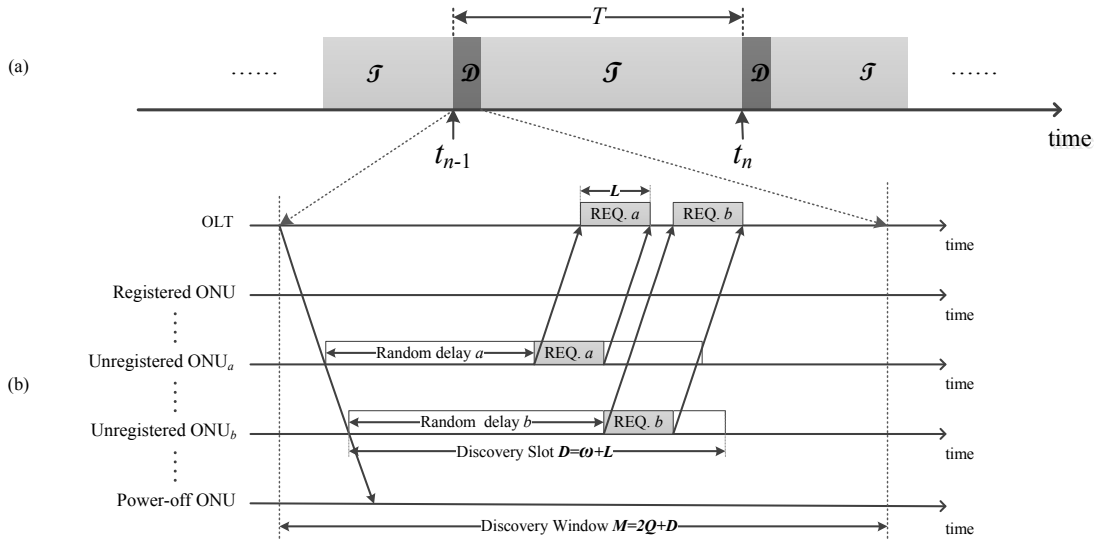}
\caption{Periodical registration processes of EPON.}\label{fig2}
\end{figure*}

In the registration protocol defined in the IEEE 802.3av standard, the maximum waiting time \cite{11Bhatia} is the only adjustable parameter in the practical operation of the network. If the maximum waiting time is set too short, excessive collisions may lead to many reattempts accumulated by unsuccessful ONUs in the registration processes. Especially in the scenario of FTTH, a large number of ONUs are geometrically clustered in a small residential area \cite{16Vaughn}, as shown in Fig. \ref{fig1}, and each ONU may frequently turn on and off for the sake of power savings. In this case, the system may become unstable and a bounded registration delay cannot be guaranteed for each ONU. On the other hand, expanding the maximum waiting time of the registration process inevitably reduces the bandwidth available for the upstream transmission of normal access service. It is therefore essential to investigate the stability and delay of the registration protocol. Some of the work that focused on the throughput analysis of the EPON registration protocol was previously reported in \cite{9cui2012throughput,10Bjelica,11Bhatia,12Hajduczenia,13Bhatia}, but the system issues on stability and delay remain open.

In this paper, we make the first attempt to analyze the stability and delay of the EPON registration protocol. In a discovery window, an ONU may be in one of the following states: registered state, unregistered state, or power-off state. Our analysis is based on the discrete-time Markov chain that describes the state transitions of a tagged ONU. The aim of the analysis is twofold: determine the stable region of the maximum waiting time and estimate the bounded delay of the registration process in the stable region. Our results demonstrate that there is a stable region in the steady-state for the maximum waiting time. In this stable region, we show that the throughput and delay of the registration protocol can be guaranteed, and the delay is a function of the maximum waiting time. Furthermore, our results reveal the fact that it is unnecessary to arbitrarily enlarge the maximum waiting time as long as it is in the stable region because the improvement of registration delay is marginal, while the reduction of registration efficiency can be quite significant.

The remainder of this paper is organized as follows. In Section \ref{section2}, we describe the registration process of EPON, and establish the discrete-time Markov chain to delineate the registration processes under the assumption of FTTH application. The main result is the derivation of the fundamental equation of EPON that characterizes the throughput of the system. In Section \ref{section3}, we study the stability of the registration protocol, and specify the stable region of the maximum waiting time for the practical implementation of EPON. In Section \ref{section4}, we investigate the average registration delay experienced by an ONU attempting to register when the system is operating in the stable region. Section \ref{conclusion} provides a conclusion of this paper.

\section{Markov Model of Registration Processes}\label{section2}
\subsection{Registration Processes}\label{subsection2A}
The EPON registration processes, as defined in the IEEE 802.3av standard and illustrated in Fig. \ref{fig2}(a), alternate between a discovery window (marked by $\mathcal{D}$ in Fig. \ref{fig2}(a)) and a normal transmission window (marked by $\mathcal{T}$ in Fig. \ref{fig2}(a)) over time. The size of a transmission window is usually much larger than that of a discovery window. In each transmission window, the network provides broadband access services for the online ONUs based on a predetermined scheduling of OLT. The normal access service is periodically interrupted by a discovery window, during which unregistered ONUs can make registration requests to the OLT if they attempt to access the network. The time interval between the start points of two consecutive discovery windows is referred to as the \emph{registration cycle}.

As plotted in Fig. \ref{fig2}(b), the OLT initiates a discovery window by broadcasting a discovery GATE message, which delivers the start time of the discovery slot and its length $D$ to all ONUs \cite{1kramer2005}. After a one-trip propagation time, the GATE message arrives at the ONU. Let $Q$ be the maximal one-trip propagation time from the OLT to the farthest ONU. In practice, $Q$ is about 100 $\mu s$ \cite{9cui2012throughput,10Bjelica,11Bhatia}. When an ONU receives the GATE message, it may be in one of the following three states: registered online state, power-off state, or registration state. The ONU ignores the GATE message if it is already registered in the online state or in the power-off state. The ONU is in the registration state only when the ONU is powered on but still invisible to the OLT.

Upon receiving the GATE message, each ONU in the registration state sends back a request message REQ of length $L<3 \mu s$ \cite{1kramer2005,9cui2012throughput,10Bjelica,11Bhatia} to the OLT after waiting for a random delay in the discovery slot. According to the IEEE 802.3av standard, the random delay is uniformly chosen from the interval $[0,\omega]$, where $\omega=D-L$, and is called the \emph{maximum waiting time}\cite{11Bhatia}. The REQ message arrives at the OLT after another one-trip propagation time. The discovery window should be large enough such that the OLT can receive all REQs. Therefore, the discovery window size is $M=2Q+D$, the sum of the maximal round-trip propagation time and the discovery slot size.

A collision occurs if two or more REQs overlap in time when they arrive at the OLT. All REQs involved in the collision are ruined and the corresponding registrations are void. The ONU that fails to register remains invisible to the OLT, and it tries to register again in the next discovery window. According to the registration process defined in the IEEE 802.3av standard, a discrete-time Markov chain that characterizes the state transitions of a tagged ONU is described in the rest of this section.

\subsection{Discrete-time Markov Chain}\label{subsection2B}
The registration processes described above can be modeled as a discrete-time Markov chain. The application of the EPON system under consideration is FTTH, which is regarded as the mainstream of broadband access networks \cite{1kramer2005,15Hutcheson}. In the scenario of FTTH, it is estimated that the number of ONUs in an EPON can reach as many as 512 in the near future \cite{7chan2010remote,8hajduczenia2010ieee}. These ONUs are usually clustered in a small housing district \cite{16Vaughn}, such that their distances to the OLT are almost the same. In addition, each ONU may frequently turn on and off to save power. Considering these points, the Markov chain is established under the following assumptions:
\begin{itemize}
  \item [1)] The ONUs are clustered in a small area, and the propagation delay from each ONU to the OLT is a small constant that our analysis ignores;
  \item [2)] The behaviors of ONUs are statistically identical in steady state, and each unregistered ONU makes an independent registration request in a discovery window;
  \item [3)] The online holding time of each registered ONU, denoted by $t_A$, is a negative exponential random variable with mean $\tau_A=E[t_A]$;
  \item [4)] The power-off holding time of each idle ONU, denoted by $t_F$, is also a negative exponential random variable with mean $\tau_F=E[t_F]$;
  \item [5)] The registration cycle time $T$ between the starting points of two consecutive discovery windows, as illustrated in Fig. \ref{fig2}(a), is assumed to be a constant.
\end{itemize}

The parameters employed in our stability and delay analysis are listed as follows for easy reference:
\begin{description}
  \item [$N$:] population of ONUs in the EPON;
  \item [$\omega$:] maximum waiting time of the discovery window;
  \item [$L$:] length of a registration request REQ message;
  \item [$G$:] aggregate traffic in a discovery window, defined as the average number of ONUs in state $R$ at the beginning of a discovery window;
  \item [$d$:] registration delay of each ONU when the system is stable;
  \item [$h$:] \emph{attempt probability}, the probability that an ONU not in the registration state will attempt to register in the next discovery window;
  \item [$p_{suc}$:] probability that an ONU attempting to register succeeds in a discovery window;
  \item [$\lambda_{out}$:] registration throughput per registration cycle in the stationary state.
\end{description}
%$N$: population of ONUs in the EPON;\\
%$\omega$: maximum waiting time of the discovery window;\\
%$L$: length of a registration request REQ message;\\
%$G$: aggregate traffic in a discovery window, defined as the average number of ONUs in state $R$ at the beginning of a discovery window;\\
%$d$: registration delay of each ONU when the system is stable;\\
%$h$: \emph{attempt probability}, the probability that an ONU not in the registration state will attempt to register in the next discovery window;\\
%$p_{suc}$: probability that an ONU attempting to register succeeds in a discovery window.

\begin{figure}[t]
\centering
\includegraphics[width=0.28\textwidth]{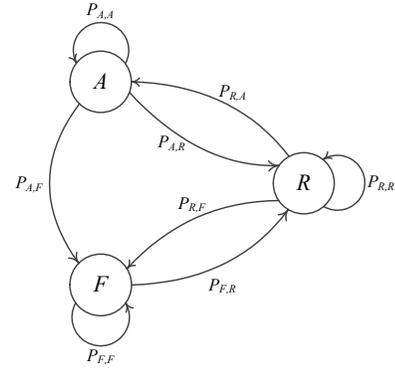}
\caption{Discrete-time Markov chain of an ONUs in EPON.}\label{fig3}
\end{figure}

Consider the state of a tagged ONU at the beginning of each discovery window. The state transitions of this ONU can be portrayed by the discrete-time Markov chain shown in Fig. \ref{fig3}. The three states $A$, $F$, and $R$ of the Markov chain represent the registered online state, power-off state, and registration state, respectively, of the ONU under consideration. From the state transition diagram, we know that the limiting probabilities $\pi_A$, $\pi_F$, and $\pi_R$ satisfy the following set of equations:
\begin{equation}\label{Eq1}
    \begin{cases}
        \pi_A (1-P_{A,A}) =  \pi_R P_{R,A} \\
        \pi_F (1-P_{F,F}) =  \pi_A P_{A,F}+\pi_R P_{R,F} \\
        \pi_R (1-P_{R,R}) =  \pi_A P_{A,R}+\pi_F P_{F,R}
    \end{cases},
\end{equation}
where $P_{x,y}$ is the transition probability for state $x$, $y\in\{A,F,R\}$.

The transition probabilities of the Markov chain can be derived from the above set of assumptions.  Let $t_n (n=1,2,\cdots)$ denote the starting point of the $n$th discovery window. Suppose that the tagged ONU is in state $F$, the power-off state, at time $t_n$. It will participate in the registration process in the next discovery window if it is turned on before time $t_{n+1}$, or equivalently, the power-off holding time $t_F$ is less than the cycle time $T$. It follows from assumption 4 above that the transition probability from state $F$ to state $R$ is given by:
\begin{equation}\label{Eq2}
  P_{F,R}=Pr\{t_F<T\}=1-e^{-\frac{T}{\tau_F}}.
\end{equation}
Since the ONU cannot directly move from the power-off state $F$ at time $t_n$ to the online state $A$ at time $t_{n+1}$ without registration, as plotted in Fig. \ref{fig3}, the probability that the ONU remains in state $F$ at time $t_{n+1}$ is given by:
\begin{equation}\label{Eq3}
  P_{F,F}=1-P_{F,R}=e^{-\frac{T}{\tau_F}}.
\end{equation}

\begin{figure*}[t]
\centering
\includegraphics[width=0.60\textwidth]{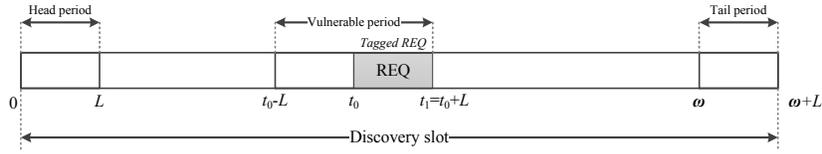}
\caption{A successful registration message.}\label{fig4}
\end{figure*}

Suppose the ONU is in the online state $A$ at time $t_n$. It could move to state $R$ at time $t_{n+1}$ if it is turned off and then turned on again before time $t_{n+1}$, which implies $t_A+t_F<T$. Thus, according to assumptions 3 and 4, we can calculate the transition probability $P_{A,R}$ from state $A$ to state $R$ as follows:
\begin{align}\label{Eq4}
  P_{A,R} &= Pr\{t_A+t_F<T\} \nonumber\\
          &= Pr\{t_A<T\}  Pr\{t_A+t_F<T|t_A<T\}\nonumber\\
          &=\bigg{(}1-e^{-\frac{T}{\tau_A}}\bigg{)} p_{rer},
\end{align}
where $p_{rer}$ is the probability that the ONU will be involved in the registration process in the next discovery window given that it is turned off before $t_{n+1}$. Again, from assumptions 3 and 4 above, we have:
\begin{align}\label{Eq5}
  p_{rer} &= Pr\{t_A+t_F<T|t_A<T\} \nonumber\\
          &= \frac{\tau_F \bigg{(}1-e^{-\frac{T}{\tau_F}}\bigg{)}-\tau_A \bigg{(}1-e^{-\frac{T}{\tau_A}}\bigg{)}}{(\tau_F-\tau_A)\bigg{(}1-e^{-\frac{T}{\tau_A}}\bigg{)}}.
\end{align}

On the other hand, the ONU in online state $A$ at time $t_n$ could move to state $F$ at time $t_{n+1}$ if it is turned off but not turned on again before time $t_{n+1}$, which implies that $t_A<T$ while $t_A+t_F>T$. Therefore, according to assumptions 3 and 4, the transition probability $P_{A,F}$ from state $A$ to state $F$ is given as follows:
\begin{align}\label{Eq6}
  P_{A,F} &= Pr\{(t_A<T) , (t_A+t_F>T)\} \nonumber\\
          &= Pr\{t_A<T\}\times Pr\{t_A+t_F>T|t_A<T\}\nonumber\\
          &= \bigg{(}1-e^{-\frac{T}{\tau_A}}\bigg{)}(1-p_{rer}).
\end{align}
From (\ref{Eq4})-(\ref{Eq6}), we have:
\begin{equation}\label{Eq7}
  P_{A,A}=1-P_{A,R}-P_{A,F}=e^{-\frac{T}{\tau_A}}.
\end{equation}

Suppose that the ONU is in the registration state $R$ at time $t_n$. It could be in the online state $A$ at time $t_{n+1}$ if it is successfully registered in the $n$th discovery window and not turned off before time $t_{n+1}$. Therefore, the transition probability $P_{R,A}$ from state $R$ to state $A$ is given as follows:
\begin{align}\label{Eq8}
  P_{R,A} &= Pr\{(\text{Registration is successful}) , (t_A \geq T)\} \nonumber\\
          &= Pr\{\text{Registration is successful}\} \times Pr\{t_A \geq T\} \nonumber\\
          &= p_{suc} \times e^{-\frac{T}{\tau_A}},
\end{align}
where $p_{suc}$ is the success probability of a registration request. In assumption 2 above, we assume that each unregistered ONU makes an independent registration request in a discovery window. For a sufficiently large population size $N$, the mean number of ONUs involved in each registration process is given by the following definition:
\begin{equation}\label{Eq9}
  G=N\pi_R.
\end{equation}
The following lemma provides the derivation of $p_{suc}$,  which is similar to that of the pure Aloha system given in \cite{21Abramson}.
\begin{lemma}\label{lemma1}
  The success probability of a registration request in a discovery window is given by:
    \begin{equation}\label{Eq10}
            p_{suc}=e^{-\frac{2L}{\omega}G}.
    \end{equation}
\end{lemma}
\begin{IEEEproof}
Suppose that a tagged REQ (registration request message) of length $L$ starts at time $t_0$ and ends at time $t_1=t_0+L$ in the discovery slot elapsed from time $t=0$ to time $t=\omega+L$, as shown in Fig. \ref{fig4}. This registration request succeeds if no other request messages start in the vulnerable period. Considering the boundary of discovery slot, there are three kinds of vulnerable period described as follows:
\begin{itemize}
  \item [1)] If $L \leq t_0 \leq \omega -L$, then the vulnerable period is from time $t_0-L$ to $t_0+L$;
  \item [2)] If $0 \leq t_0 < L$, when the REQ message overlaps into the head period, then the vulnerable period is from time $0$ to $t_0+L$;
  \item [3)] If $\omega-L<t_0 \leq \omega $, 	when the REQ message overlaps into the tail period, then the vulnerable period is from time $t_0-L$, to $\omega$.
\end{itemize}
%1. If $L \leq t_0 \leq \omega -L$, 	then the vulnerable period is from time $t_0-L$ to $t_0+L$;\\
%2. If $0 \leq t_0 < L$, when the REQ message overlaps into the head period, then the vulnerable period is from time $0$ to $t_0+L$;\\
%3. If $\omega-L<t_0 \leq \omega $, 	when the REQ message overlaps into the tail period, then the vulnerable period is from time $t_0-L$, to $\omega$.

Since the maximum waiting time $\omega$ is much larger than the message length $L$, and the probability that a REQ message overlaps into head or tail period is negligible, we can ignore cases 2 and 3 above in the following derivation. Given that the starting point of a message is uniformly distributed in the time interval $[0,\omega]$, then the probability $q$ that an arbitrary ONU will not interfere with this tagged registration message is given by:
\begin{align}\label{Eq11}
 q =& Pr\{\text{ONU not in state $R$}\} +Pr\{(\text{ONU in state $R$}),  \nonumber \\
    & (\text{request message not starting in vulnerable period})\}  \nonumber \\
   \cong & (1- \pi_R) + \pi_R \Big{(} 1- \frac{2L}{\omega}\Big{)} \nonumber \\
   =& 1- \frac{2L}{\omega} \pi_R.
\end{align}
According to assumption 2, we obtain:
\begin{align*}
  p_{suc}&= q^{N-1} \cong \Big{(} 1-\frac{2L}{\omega}\pi_R \Big{)}^{N-1} \\
         &=\bigg{[} 1-\frac{1}{N-1}\Big{(}\frac{2LN\pi_R}{\omega}\Big{)}\frac{N-1}{N} \bigg{]}^{N-1}.
\end{align*}
For $N\gg1$, we have:
\begin{equation*}
  p_{suc} \cong \bigg{[} 1-\frac{1}{N-1}\Big{(}\frac{2L}{\omega}G \Big{)} \bigg{]}^{N-1} \cong e^{-\frac{2L}{\omega}G}.
\end{equation*}
\end{IEEEproof}
%If $\omega \leq L$, once collision occurs, all the ONUs will eventually be caught in the registration state R and none of them has any chances to escape. If the length of the discovery slot is as short as several REQ messages, it is highly probable that the registration protocol should collapse. Hence, in the following, we will only focus on the situation where $\omega \gg L$.

%Because the vulnerable period of the tagged REQ varies with its starting point $t_0$, there are the following 3 cases: if $0 \leq t_0 < L$, the vulnerable period is $[0,t_0+L]$; if $L \leq t_0 \leq \omega -L$, the vulnerable period is $[t_0-L,t_0+L]$; if $\omega-L<t_0 \leq \omega $, the vulnerable period is $[t_0-L,\omega]$.

%\begin{align*}
% q =& Pr\{\text{ONU not in state $R$}\} +Pr\{(\text{ONU in state $R$}),  \\
%    & (\text{request message not starting in vulnerable period})\}  \\
%    =& (1- \pi_R) + \pi_R \Bigg {[}\overset{L}{\underset{0}{\int}} \Big{(} 1- \frac{t_0+L}{\omega}\Big{)}\frac{1}{\omega}dt_0   \\
 %    & + \overset{\omega-L}{\underset{L}{\int}}\Big{(} 1- \frac{2L}{\omega}\Big{)}\frac{1}{\omega}dt_0 + \overset{\omega}{\underset{\omega-L}{\int}} \Big{(} 1- \frac{\omega -t_0+L}{\omega}\Big{)}\frac{1}{\omega}dt_0 \Bigg{]}.
%\end{align*}
%Since $\omega \gg L$, we can have:
%\begin{align}\label{Eq11}
%  q \cong & (1- \pi_R) + \pi_R \overset{\omega}{\underset{0}{\int}}\Big{(} 1- \frac{2L}{\omega}\Big{)}\frac{1}{\omega}dt_0 \nonumber \\
%      =   & (1-\pi_R)+\pi_R \Big{(} 1-\frac{2L}{\omega} \Big{)}\nonumber\\
%      =   & 1-\frac{2L}{\omega}\pi_R.
%\end{align}

The above derivation is consistent with the result that we obtained from the throughput analysis of EPON reported in \cite{9cui2012throughput}. It is possible that the ONU could also move into the power-off state $F$ if it is successfully registered in the $n$th discovery window but turned off before time $t_{n+1}$. Consequently, the transition probability $P_{R,F}$ from state $R$ to state $F$ can be determined as follows:
\begin{align}\label{Eq12}
  P_{R,F} =& Pr\{ (\text{Registration is successful}), (t_A<T), (t_A+ \nonumber\\
           & t_F>T)\} \nonumber\\
          =& p_{suc} \times Pr\{(t_A<T) , (t_A+t_F>T) \} \nonumber \\
          =& p_{suc} \times Pr\{t_A<T\} \times Pr\{t_A+t_F>T|t_A<T \} \nonumber \\
          =& p_{suc} \times  \bigg{(}1-e^{-\frac{T}{\tau_A}} \bigg{)} \times (1-p_{rer}).
\end{align}
Again, from (\ref{Eq8}) and (\ref{Eq12}), we obtain the transition probability $P_{R,R}$ that the ONU remains in state $R$ as follows:
\begin{align}\label{Eq13}
  P_{R,R} &= 1-P_{R,A}-P_{R,F} \nonumber \\
          &= 1-p_{suc}+p_{suc} \times \bigg{(} 1-e^{-\frac{T}{\tau_A}} \bigg{)} \times p_{rer}.
\end{align}

In practice, the cycle time $T$ is on the order of hundreds of milliseconds \cite{1kramer2005,17Hajduczenia}, and it is much smaller than $\tau_A$ and $\tau_F$, because users are not likely to turn on and off the ONUs very frequently. Based on this practical condition, we obtain the following approximations, which are helpful to make the computation of our model more tractable.
\begin{lemma} \label{lemma2}
  If $\tau_A$, $\tau_F \gg T$, then we have:\\
A1) $p_{rer}\approx 0;$\\
A2) $ \frac{\pi_A}{\pi_F} \cong \frac{\tau_A}{\tau_F}.$
\end{lemma}
\begin{IEEEproof}
  Ignoring the higher order terms in the Taylor's series expansion $e^{-x}\approx 1-x+o(x)$ for $|x| \ll 1$, we obtain A1 from (\ref{Eq5}) as follows:
\begin{equation*}
  p_{rer} \approx \frac{\tau_F \bigg{[}1-\Big{(}1-\frac{T}{\tau_F}\Big{)}\bigg{]}-\tau_A \bigg{[}1-\Big{(} 1-\frac{T}{\tau_A}\Big{)}\bigg{]}}{(\tau_F-\tau_A)\bigg{(}1-e^{-\frac{T}{\tau_A}}\bigg{)}}=0.
\end{equation*}
From the first and second balance equations of (\ref{Eq1}) and the transition probabilities given by (\ref{Eq3}), (\ref{Eq6})-(\ref{Eq8}), and (\ref{Eq12}), we have:
\begin{align*}
  \frac{\pi_A}{\pi_F} =& \frac{(1-P_{F,F})P_{R,A}}{P_{A,F}P_{R,A}+(1-P_{A,A})P_{R,F}} \\
                      =& \frac{\bigg{(} 1-e^{-\frac{T}{\tau_F}}\bigg{)} \times e^{-\frac{T}{\tau_A}}}{\bigg{(} 1-e^{-\frac{T}{\tau_A}}\bigg{)} \times (1-p_{rer})}.
\end{align*}
Similarly, we obtain A2 as follows:
\begin{equation*}
  \frac{\pi_A}{\pi_F} \cong \frac{\bigg{[} 1- \Big{(} 1-\frac{T}{\tau_F} \Big{)}\bigg{]} \times \Big{(} 1-\frac{T}{\tau_A} \Big{)}} {\bigg{[} 1-\Big{(}1-\frac{T}{\tau_A} \Big{)} \bigg{]}} = \frac{\tau_A \Big{(} 1-\frac{T}{\tau_A}\Big{)}}{\tau_F} \cong \frac{\tau_A}{\tau_F}.
\end{equation*}
\end{IEEEproof}

Given the transition probabilities (\ref{Eq2})-(\ref{Eq8}) and (\ref{Eq12})-(\ref{Eq13}) in the steady-state, the set of equations (\ref{Eq1}) can be simultaneously solved with the characteristic equation of throughput given in the following theorem.
\begin{theorem}\label{theorem1}
  In the steady-state of the EPON registration process, the limiting probability $\pi_R$ of the Markov chain satisfies the following characteristic equation:
\begin{equation}\label{Eq14}
  (1-\pi_R)h=\pi_R e^{-\frac{2LN}{\omega}\pi_R},
\end{equation}
where $h$ is a constant approximately given by:
\begin{equation}\label{Eq15}
  h \cong \frac{T}{\tau_A+\tau_F}.
\end{equation}
\end{theorem}
\begin{IEEEproof}
  In the steady-state of the EPON registration process, from the first balance equation of (\ref{Eq1}), we obtain:
\begin{equation}\label{Eq16}
  \pi_A=\frac{\pi_R P_{R,A}}{1-P_{A,A}}.
\end{equation}
Clearly, we have $\pi_A+\pi_F+\pi_R=1$, because each ONU must be in one of the three states at the beginning of any discovery window. Substituting (\ref{Eq16}) into the second balance equation of (\ref{Eq1}), we obtain:
\begin{equation}\label{Eq17}
  \Big{(}1-\pi_R-\frac{\pi_R P_{R,A}}{1-P_{A,A}} \Big{)}(1-P_{F,F})=\frac{\pi_R P_{R,A}}{1-P_{A,A}} P_{A,F}+\pi_R P_{R,F}.
\end{equation}
Now, we can derive (\ref{Eq14}) from (\ref{Eq17}) together with the transition probabilities given by (\ref{Eq3}), (\ref{Eq6})-(\ref{Eq8}), and (\ref{Eq12}), where $h$ is a constant given as follows:
\begin{equation*}
  h=\frac{\bigg{(} 1-e^{-\frac{T}{\tau_A}} \bigg{)} \bigg{(} 1-e^{-\frac{T}{\tau_F}} \bigg{)}}{1-e^{-\big{(}\frac{T}{\tau_A}+\frac{T}{\tau_F} \big{)} }-p_{rer} \bigg{(} 1-e^{-\frac{T}{\tau_A}} \bigg{)}}.
\end{equation*}
The approximation of $h$ given by (\ref{Eq15}) can be readily obtained from A1 of Lemma \ref{lemma2}.
\end{IEEEproof}

\begin{figure}[b]
\centering
\includegraphics[width=0.40\textwidth]{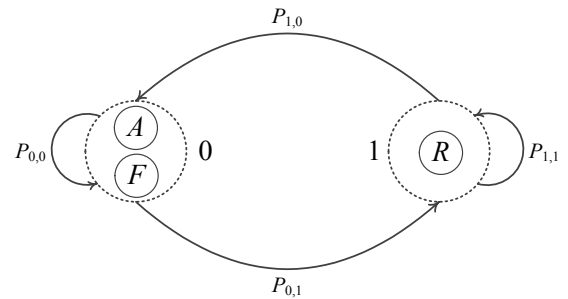}
\caption{The integrated two-state Markov chain.}\label{fig5}
\end{figure}

The above theorem gives the fundamental equation (\ref{Eq14}) of EPON that characterizes the throughput of the system. Multiplying both sides of (\ref{Eq14}) by $N$, we obtain:
\begin{equation}\label{Eq18}
  N(1-\pi_R)h=N\pi_R e^{-\frac{2LN}{\omega}\pi_R}.
\end{equation}
The right-hand side of (\ref{Eq18}) is equal to the average number of ONUs successfully registered per registration cycle, $N\pi_R p_{suc}$, or the \emph{departure rate} of the registration process, denoted by $\lambda_{out}$. On the left-hand side, $N(1-\pi_R)$ is the average number of ONUs not in the registration state $R$. We want to demonstrate that the characteristic equation (\ref{Eq14}) is the balance equation between the \emph{arrival rate} $N(1-\pi_R)h$ and the departure rate of registration requests, and the parameter $h$ is the \emph{attempt probability}, defined as the probability that an ONU not in the registration state will attempt to register in the next discovery window. This point can be elaborated by the integrated two-state Markov chain shown in Fig.\ref{fig5}.

\begin{figure*}[t]
\centering
\mbox{
    \subfigure[]{\label{fig6a}
    \begin{minipage}[c]{0.43\textwidth}
    \centering
    \includegraphics[width=1\textwidth]{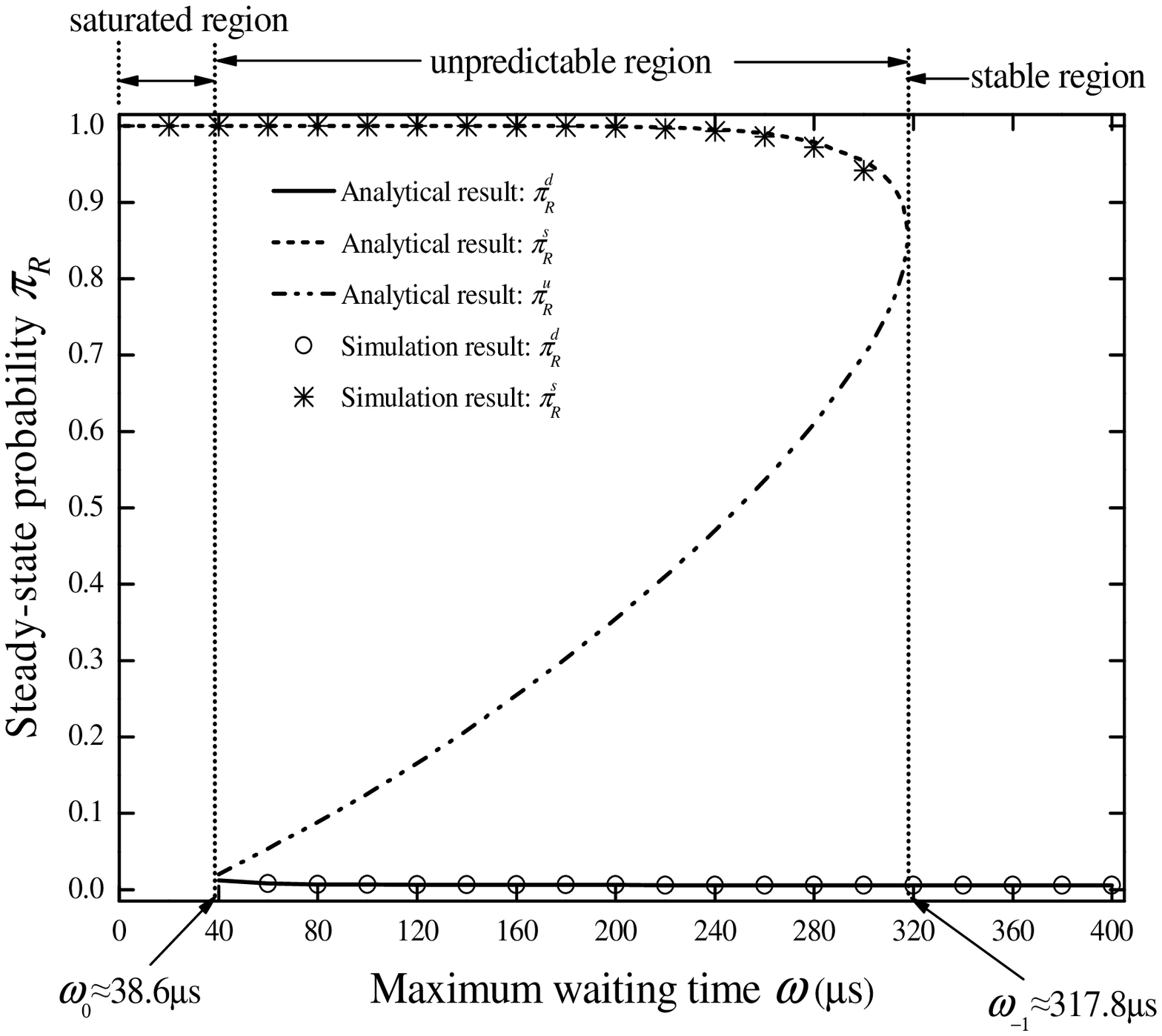}
    %\vspace{0.001\textwidth}
    \end{minipage}%
    }%注意这个”%”绝对不能省，可以试试不打%的效果

\hspace{0.05\textwidth}

    \subfigure[]{\label{fig6b}
    \begin{minipage}[c]{0.43\textwidth}
    \centering
    \includegraphics[width=1\textwidth]{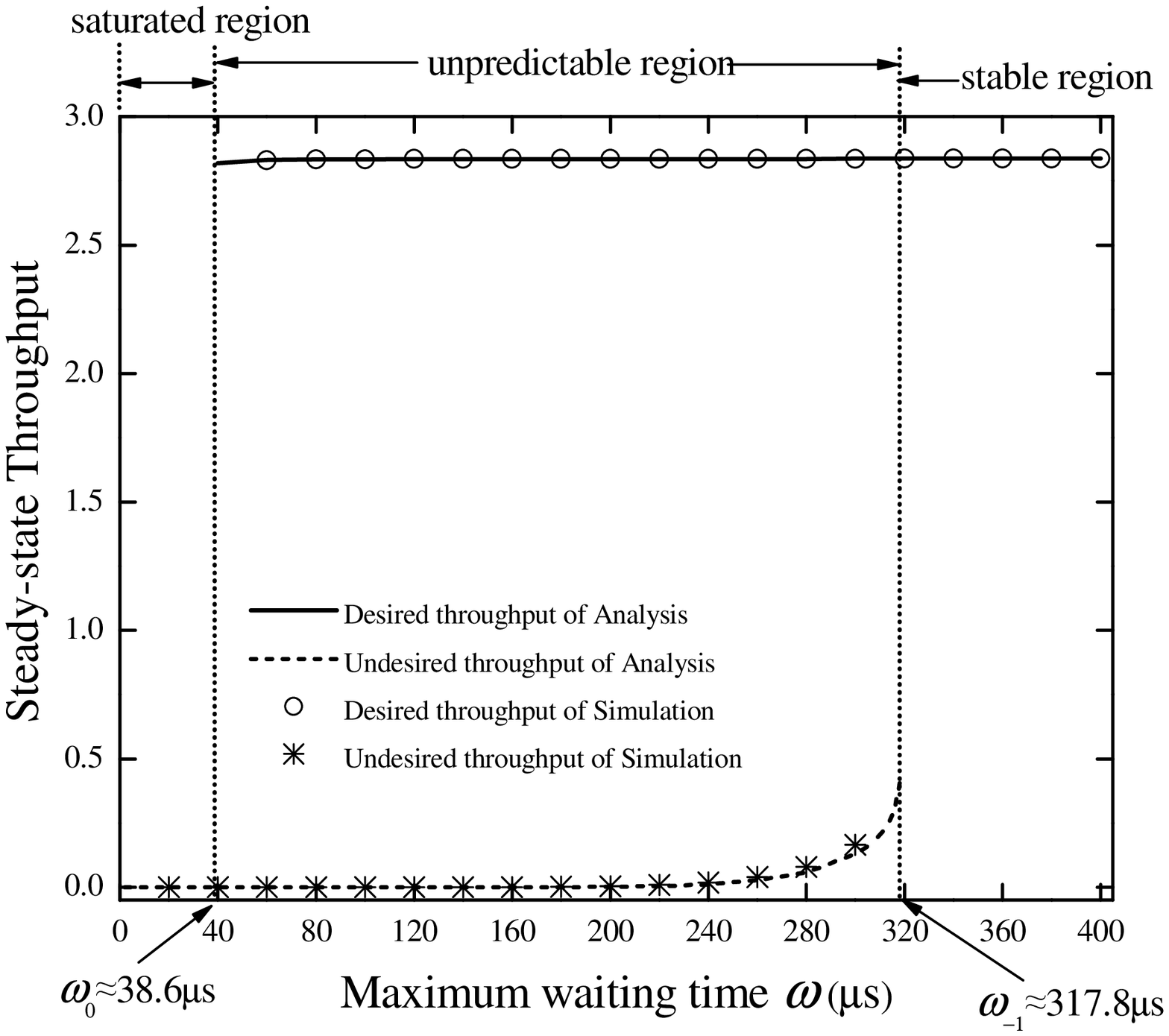}
    %\vspace{0.001\textwidth}
    \end{minipage}%
    }%注意这个”%”绝对不能省，可以试试不打%的效果
}
\caption{Different regions of $\omega$ for (a) the solutions of the characteristic equation (14)and (b) the steady-state throughput.}\label{fig6}
\end{figure*}

In the steady-state, an ONU is either in the registration state $R$ with probability $\pi_R$, or in the other two non-registration states, the online state $A$ and the power-off state $F$, with probability $\pi_A+\pi_F=1-\pi_R$. The discrete-time Markov chain shown in Fig. \ref{fig3} is thus condensed into the two-state \{0,1\} Markov chain displayed in Fig. \ref{fig5}, where the registration state 1 corresponds to state $R$, while the non-registration state 0 represents the integration of state $A$ and state $F$. Let random variable $\chi_n$ denote the state of an ONU at the time $t_n$. The transition probability from the non-registration state 0 to the registration state 1 is then defined by:
\begin{align}\label{Eq19}
  P_{0,1} = & Pr\{\chi_{n+1}=1|\chi_n=0\} \nonumber \\
          = & Pr\{\chi_{n+1}=1|\text{state $A$ at $t_n$}\}Pr\{\text{state $A$ at $t_n$}|\chi_n=0\} \nonumber \\
            & +Pr\{\chi_{n+1}=1|\text{state $F$ at $t_n$}\} \times \nonumber \\
            & ~~~Pr\{\text{state $F$ at $t_n$}|\chi_n=0\} \nonumber \\
          = & P_{A,R} \frac{\pi_A}{\pi_A+\pi_F}+P_{F,R} \frac{\pi_F}{\pi_A+\pi_F}.
\end{align}
Also, from (\ref{Eq19}) and the definition of limiting probability $\pi_A+\pi_F=1-\pi_R$, we have:
\begin{equation}\label{Eq20}
  (1-\pi_R)P_{0,1}=\pi_A P_{A,R}+\pi_F P_{F,R}.
\end{equation}
From the third balance equation of (\ref{Eq1}) and the transition probability given by (\ref{Eq13}), we have:
\begin{align}\label{Eq21}
  (1-\pi_R) P_{0,1} & = \pi_A P_{A,R}+\pi_F P_{F,R} \nonumber \\
                    & = \pi_R (1-P_{R,R}) \nonumber \\
                    & = \pi_R p_{suc} \Bigg{[}1- \bigg{(}1-e^{-\frac{T}{\tau_A}} \bigg{)} \times p_{rer} \Bigg{]} \nonumber \\
                    & \approx \pi_R p_{suc}.
\end{align}
The last step is due to the approximation A1 given in Lemma \ref{lemma2}. Substituting (\ref{Eq2}) and (\ref{Eq4}) into (\ref{Eq19}), from Lemma \ref{lemma2}, we obtain:
\begin{align}\label{Eq22}
  P_{0,1}        = & P_{A,R} \frac{\pi_A}{\pi_A+\pi_F}+P_{F,R}\frac{\pi_F}{\pi_A+\pi_F}  \nonumber \\
           \approx & \bigg{(} 1-e^{-\frac{T}{\tau_A}} \bigg{)} \frac{p_{rer} \tau_A}{\tau_A+\tau_F}+\bigg{(} 1-e^{-\frac{T}{\tau_F}} \bigg{)} \frac{\tau_F}{\tau_A+\tau_F} \nonumber \\
           \approx & \frac{T}{\tau_A+\tau_F}.
\end{align}

Comparing (\ref{Eq14}) and (\ref{Eq21}), the above expression (\ref{Eq22}) confirms the fact that the attempt probability $h$ given by (\ref{Eq15}) is indeed the transition probability $P_{0,1}$ that an ONU not in the registration state will attempt to register in the next discovery window, and $N(1-\pi_R)h$ is the traffic that arrives per registration cycle.

\begin{figure*}[t]
\centering
\mbox{
    \subfigure[Saturated region:~$\omega < \omega_0$]{\label{fig7a}
    \begin{minipage}[c]{0.32\textwidth}
    \centering
    \includegraphics[width=1\textwidth]{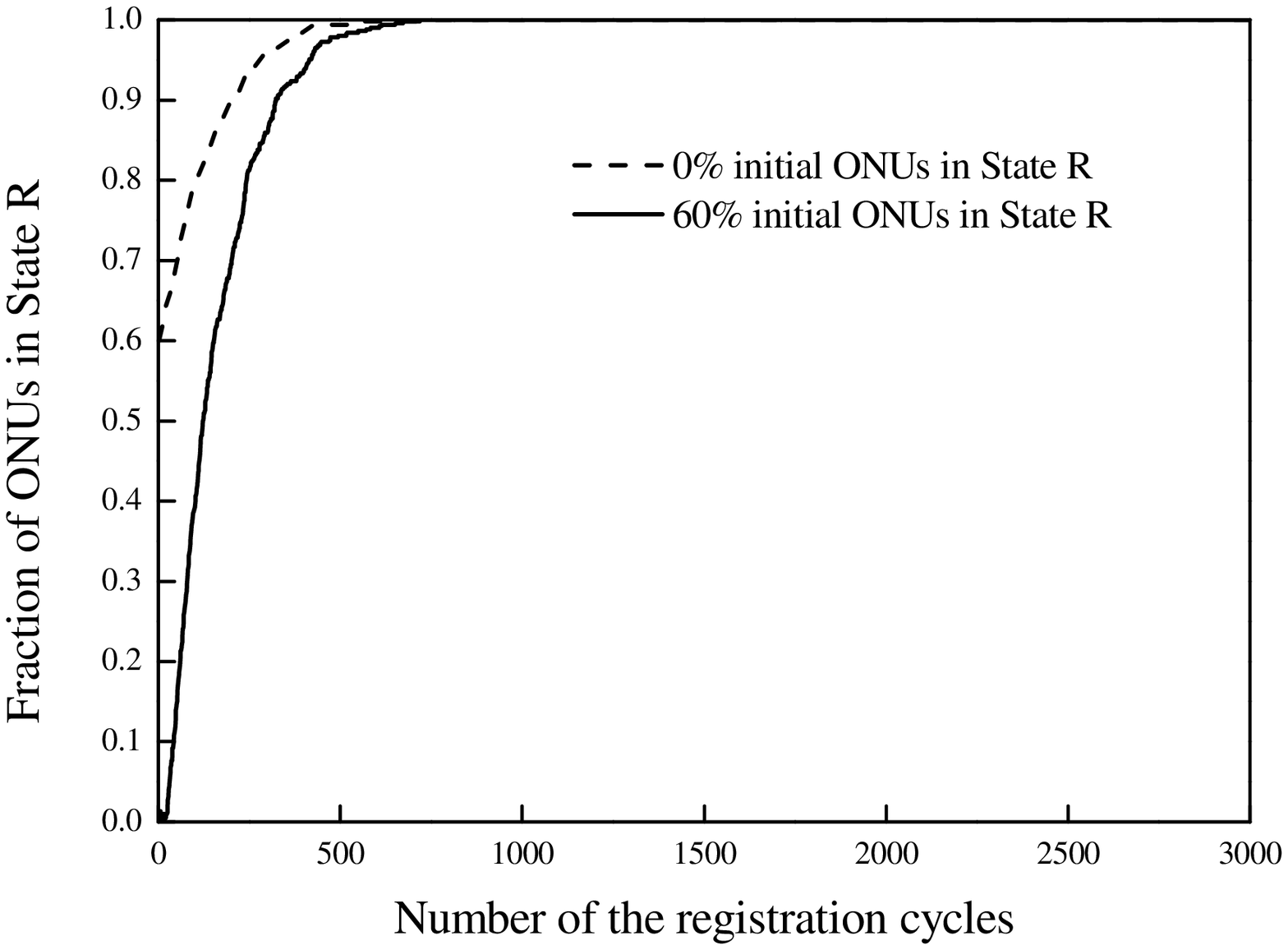}
    \vspace{0.001\textwidth}
    \end{minipage}%
    }%注意这个”%”绝对不能省，可以试试不打%的效果

    \subfigure[Unpredictable region:~$\omega_0 < \omega < \omega_{-1}$]{\label{fig7b}
    \begin{minipage}[c]{0.32\textwidth}
    \centering
    \includegraphics[width=1\textwidth]{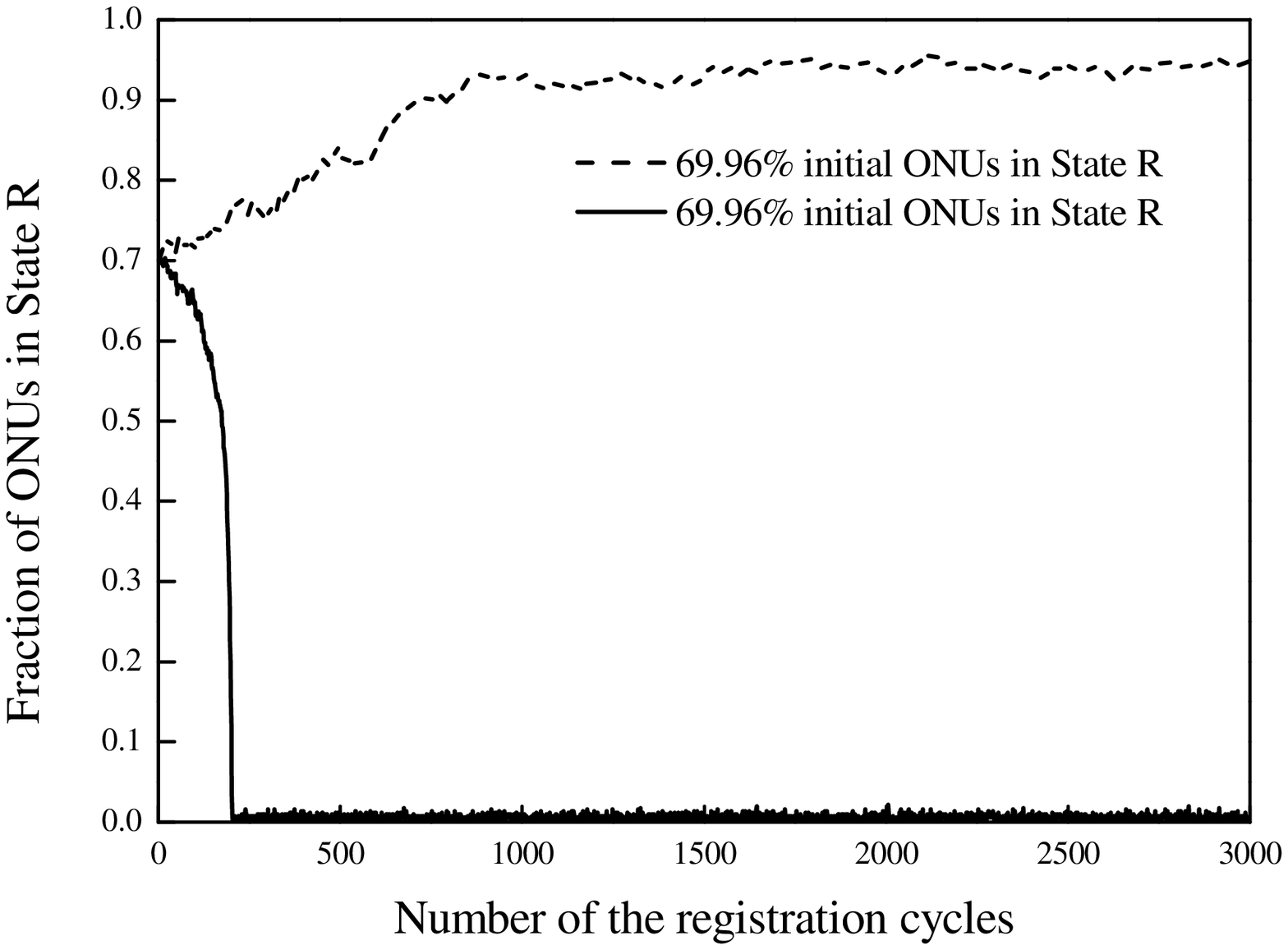}
    \vspace{0.001\textwidth}
    \end{minipage}%
    }%注意这个”%”绝对不能省，可以试试不打%的效果

    \subfigure[Stable region:~$\omega > \omega_{-1}$]{\label{fig7c}
    \begin{minipage}[c]{0.32\textwidth}
    \centering
    \includegraphics[width=1\textwidth]{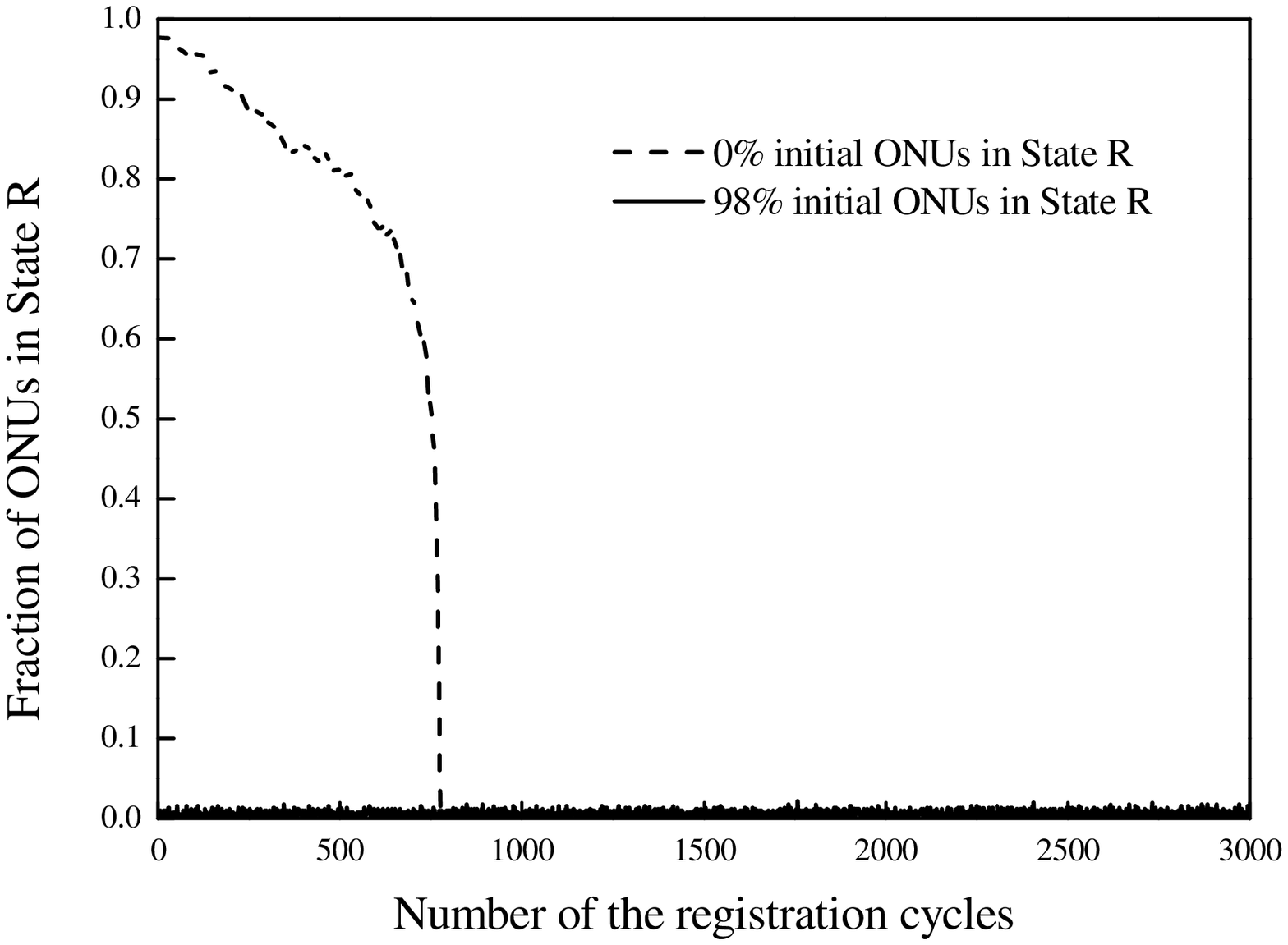}
    \vspace{0.001\textwidth}
    \end{minipage}%
    }
}
\caption{Simulation results with different maximum waiting time $\omega$}\label{fig7}
\end{figure*}

\section{Stability Conditions}\label{section3}
In this section, we explore the stability analysis of the EPON registration protocol. For a given set of EPON parameters, Theorem \ref{theorem1} states that the throughput of the protocol is determined by the maximum waiting time $\omega$ by solving equation (\ref{Eq14}) for the limiting probability $\pi_R$. Intuitively, $\omega$ has to be sufficiently large to alleviate contentions among registration requests generated by ONUs.

For instance, suppose the EPON parameters are $N=512$, $\tau_A=60s$, $\tau_F=30s$, and $T=500ms$, then the limiting probability $\pi_R$ calculated from (\ref{Eq14}) can be as small as $0.00578$ and the steady-state registration throughput is $\lambda_{out} \cong 2.83$ when $\omega=350 \mu s$, while the limiting probability $\pi_R \cong 1$ and $\lambda_{out} \cong 0$ when $\omega=30 \mu s$. This example indicates that the ONU has almost no chance to succeed in the registration process if the maximum waiting time $\omega$ is too small. When the limiting probability $\pi_R$ is very close to 1, nearly all the ONUs participate in the registration process, and eventually the EPON collapses. The aim of our stability analysis is to specify the region of the maximum waiting time $\omega$, in which the limiting probability $\pi_R$ can be sufficiently small and the EPON system has a stable throughput. To facilitate the presentation, the above EPON parameters are used as a running example throughout this paper.

We show in Appendix \ref{appendixchar} that the solutions of the equation (\ref{Eq14}) for the limiting probability $\pi_R$ depend on the regions of $\omega$ specified by the following two critical parameters:
\begin{equation}\label{Eq23}
  \omega_0= -\frac{2LNW_0(\alpha)}{[1-W_0(\alpha)]^2},
\end{equation}
and
\begin{equation}\label{Eq24}
  \omega_{-1}= -\frac{2LNW_{-1}(\alpha)}{[1-W_{-1}(\alpha)]^2},
\end{equation}
where $\alpha=-eh$ is a constant, and $W_0(\alpha)$ in (\ref{Eq23}) and $W_{-1}(\alpha)$ in (\ref{Eq24}) are the two possible values of the Lambert $W$ function, which Appendix \ref{appendixW} briefly describes. These values of the Lambert $W$ function are defined in the range of $\alpha \geq -e^{-1}$ or equivalently $h \leq e^{-2}$. According to (\ref{Eq15}), this condition imposes the following implementation requirement:
\begin{equation}\label{Eq25}
  T \leq \frac{\tau_A+\tau_F}{e^2},
\end{equation}
which should normally hold in most practical EPON systems because typically we have $\tau_A \gg T $ and $\tau_F \gg T$. Later in this section, we show that a strictly stable registration protocol actually requires a stricter condition on the attempt probability $h$.

%\begin{figure}[t]
%\centering
%\includegraphics[width=0.45\textwidth]{fig6.eps}
%\caption{Different regions of $\omega$ for solutions of the characteristic equation of the registration protocol.}\label{fig6}
%\end{figure}

The limiting probability $\pi_R$ and the registration throughput $\lambda_{out}$ versus the maximum waiting time $\omega$ of the discovery window is plotted in Fig. \ref{fig6}, in which we use the EPON parameters $N=512$, $\tau_A=60s$, $\tau_F=30s$, and $T=500ms$, the same as the above running example. The number of solutions of equation (\ref{Eq14}) changes with respect to $\omega$. In Appendix \ref{appendixchar}, the characteristic equation (\ref{Eq14}) has one solution if $\omega < \omega_0$ or $\omega > \omega_{-1}$, and multiple solutions if $\omega_0 \leq \omega \leq \omega_{-1}$. The properties of these solution regions are described below.

\subsection{Saturated Region}\label{subsection3A}
In the region $\omega < \omega_0$, the equation (\ref{Eq14}) of the limiting probability $\pi_R$ has a unique solution, denoted as $\pi_R^s$. In this region, the maximum waiting time $\omega$ is so small that almost all registration requests will be ruined due to contention. Consequently, the solution $\pi_R^s$ is very close to 1, and the throughput $\lambda_{ouot} \cong 0$, as Fig. \ref{fig6} displays. Consider the extreme case when $\omega=0$. It is easy to show from (\ref{Eq14}) that the limiting probability $\pi_R^s$ approaches 1 and $\lambda_{out}$ vanishes. This phenomenon is also revealed by the simulation results displayed in Fig. \ref{fig7a}, where the two curves, respectively, correspond to 0\% and 60\% initial ONUs in state $R$ with $\omega = 38 \mu s$. In both cases, the limiting probability $\pi_R$ converges to 1. Eventually, the system will be on the verge of collapse after a finite number of registration cycles.

\subsection{Unpredictable Region}\label{subsection3B}
In the region $\omega_0 < \omega < \omega_{-1}$, the equation (\ref{Eq14}) of the limiting probability $\pi_R$ has three solutions, denoted as $\pi_R^s$, $\pi_R^d$ and $\pi_R^u$, where $\pi_R^s > \pi_R^u > \pi_R^d$, as shown in Fig. \ref{fig6a}. It is a well-known property of an ergodic Markov chain that the limiting probability must be unique \cite{18ross2006introduction,19borovkov1998ergodicity}, thus the performance of ONUs is unpredictable in this region because the system is obviously not ergodic. For the running example with $\omega = 300 \mu s$, we first obtain the solution $\pi_R^u = 69.93\%$ by solving equation (\ref{Eq14}). However, the simulation results indicate that the limiting probability $\pi_R$ can either converge to $\pi_R^s=94.97\%$ or $\pi_R^d=0.58\%$, as displayed in Fig. \ref{fig7b}, with the same 69.96\% ONUs in state $R$ initially. Therefore, the registration protocol may result in different throughputs depending on the steady-state limiting probability $\pi_R$, as demonstrated in Fig. \ref{fig6b}. For this reason, the performance of any individual ONU cannot be guaranteed in this unpredictable region.

\subsection{Stable Region}\label{subsection3C}
In the region $\omega > \omega_{-1}$, we show in Appendix \ref{appendixchar} that the equation (\ref{Eq14}), which specifies the limiting probability of $\pi_R$, has only one solution that converges to $\pi_R^d$. Accordingly, the system can achieve a stable throughput $\lambda_{out}$, as shown in Fig. \ref{fig6b}. Presumably, a stable registration protocol requires a very small limiting probability $\pi_R$. We show in the following that the attempt probability $h$ must also be kept small even if the maximum waiting time selected from the stable region $\omega > \omega_{-1}$ is large enough. Intuitively, if users behave maliciously and turn on and off ONUs too frequently, meaning that $\tau_A+\tau_F$ is too small, then the stable throughput may be achieved at the expense of a large mean delay.

\begin{figure}[t]
\centering
\includegraphics[width=0.40\textwidth,height=0.323\textwidth]{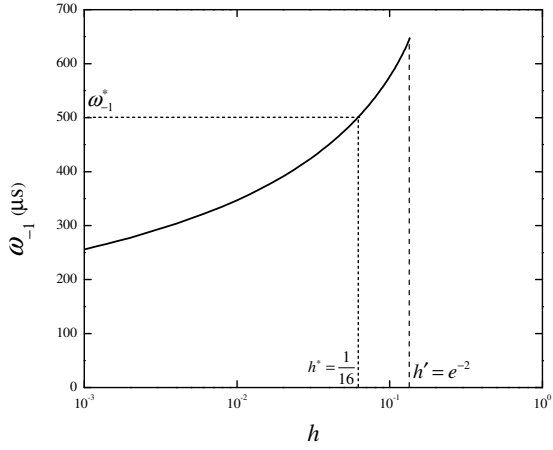}
\caption{Parameter $\omega_{-1}$ versus  attempt probability $h$.}\label{fig8_omega_h}
\end{figure}

If the system is stable, and the limiting probability $\pi_R$ converges to $\pi_R^d \ll 1$, then we should be able to approximately rewrite equation (\ref{Eq14}) as follows:
\begin{equation*}
  (1-\pi_R)h \cong \pi_R \Big{(} 1-\frac{2LN}{\omega} \pi_R \Big{)},
\end{equation*}
from which we obtain the following approximation of the limiting probability $\pi_R$:
\begin{equation}\label{Eq26Pi_R}
  \pi_R \cong \frac{1-\sqrt{1-\frac{8LNh}{(1+h)^2\omega}}}{4LN}(1+h)\omega.
\end{equation}
The maximum waiting time $\omega$ should satisfy the following condition if the above expression (\ref{Eq26Pi_R}) is valid:
\begin{equation*}
  \omega \geq \frac{8LNh}{(1+h)^2}.
\end{equation*}
Thus, the stable condition of an EPON registration protocol is formally defined as follows:
\begin{definition}\label{stable_condition}
  \emph{The EPON registration protocol is stable if the maximum waiting time $\omega$ is selected from the range $\omega > \omega_{-1}$. Moreover, the protocol is strictly stable if the attempt probability $h$ satisfies the following inequality:}
\begin{equation}\label{Eq27}
  \omega > \omega_{-1} > \frac{8LNh}{(1+h)^2}.
\end{equation}
\end{definition}
In the following corollary, the above inequality (\ref{Eq27}) induces a necessary condition for a strictly stable registration protocol.
\begin{corollary}\label{corollary1}
  A strictly stable registration protocol satisfies the following necessary condition:
\begin{equation}\label{Eq28}
  h  < \frac{1}{16}.
\end{equation}
\end{corollary}
\begin{IEEEproof}
According to (\ref{Eq24}), the stable condition requires that:
\begin{equation}\label{Eq29}
  \omega \geq \omega_{-1} = -\frac{2LNW_{-1}(\alpha)}{[1-W_{-1}(\alpha)]^2}.
\end{equation}
Thus, the following condition on the attempt probability $h>0$ ensures that both inequalities (\ref{Eq27}) and (\ref{Eq29}) can always be satisfied:
\begin{equation*}
  -\frac{2LNW_{-1}(\alpha)}{[1-W_{-1}(\alpha)]^2} > 8LNh > \frac{8LNh}{(1+h)^2},
\end{equation*}
which is equivalent to:
\begin{equation}\label{Eq30_equivalence}
  h < -\frac{W_{-1}(\alpha)}{4[1-W_{-1}(\alpha)]^2}.
\end{equation}
We know from Fig. \ref{figW} in Appendix \ref{appendixW} that $W_{-1} (\alpha) = -c^2$  for some $c \geq 1$, therefore the above inequality implies that:
\begin{equation*}
  h < \frac{c^2}{4(1+c^2)^2} = \frac{1}{4 {\Big{(} \frac{1}{c} + c \Big{)}}^2} \leq \frac{1}{16}.
\end{equation*}
\end{IEEEproof}

\begin{figure}[t]
\centering
\includegraphics[width=0.40\textwidth,height=0.323\textwidth]{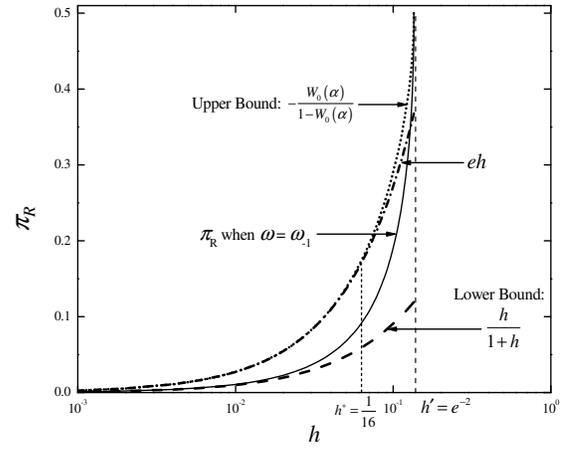}
\caption{The limiting probability $\pi_R$ versus attempt probability $h$}\label{fig9_Pi_R_h}
\end{figure}

Using the same EPON parameters of our running example, we show in Fig. \ref{fig8_omega_h} that parameter $\omega_{-1}$ increases with respect to $h$. In particular, parameter $\omega_{-1}$ skyrockets when $h>1/16$, which clearly demonstrates the necessity of the condition given by (\ref{Eq28}) for a strictly stable registration protocol.

\begin{figure*}[t]
\centering
\includegraphics[width=0.70\textwidth]{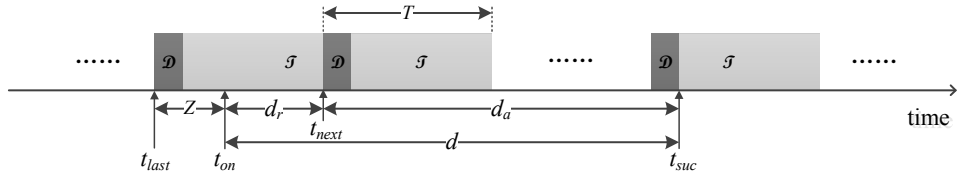}
\caption{Components of the registration delay.}\label{fig10}
\end{figure*}

We know from (\ref{Eq14}) that the arrival rate of registration requests per cycle time $T$ is $N(1-\pi_R)h$. For a given registration cycle $T$, a smaller attempt probability $h$ implies larger $\tau_A$ and $\tau_F$, meaning that the users tend to switch the ONUs on and off less frequently. Therefore, the ONUs are less likely to engage in a registration process. As a result, the arrival traffic for registration requests becomes smaller, which in turn decreases the probability of collision. Therefore, for a smaller attempt probability $h$, a smaller maximum waiting time $\omega$ is required to moderate the collisions. This point is reinforced in the following corollary.
\begin{corollary}\label{corollary2}
  Suppose the maximum waiting time is selected from the region $\omega > \omega_{-1}$. Then the unique limiting probability $\pi_R$ is bounded by:
\begin{equation}\label{Eq31}
  \frac{h}{1+h} \leq \pi_R \leq -\frac{W_0(\alpha)}{1-W_0(\alpha)} \cong eh,
\end{equation}
where the approximation of the upper bound is valid for small attempt probability $h$.
\end{corollary}
\begin{IEEEproof}
  In Appendix \ref{appendixchar}, we prove that if $\omega > \omega_{-1}$ then the equation (\ref{Eq14}), which specifies the limiting probability of $\pi_R$, has a unique solution in the interval $\big{[}0,-\frac{W_0(\alpha)}{1-W_0(\alpha)}\big{]}$. In Appendix \ref{appendixW}, we show that $W_0(\alpha)$ has the following series expansion:
\begin{equation*}
  W_0(\alpha)=\overset{\infty}{\underset{i=1}\sum}{\frac{(-i)^{i-1}}{i!}\alpha^i}=\alpha-\alpha^2+\frac{3}{2}\alpha^3-\frac{8}{3}\alpha^4+\cdots,
\end{equation*}
which is approximately equal to $\alpha=-eh$ when the attempt probability $h$ is small. That is, if the maximum waiting time $\omega$ is large enough and the attempt probability $h$ is small, i.e., $-\alpha \ll 1$, then we have:
\[\pi_R \ll -\frac{W_0(\alpha)}{1-W_0(\alpha)} \approx -W_0(\alpha) \approx -\alpha = eh. \]
When the maximum waiting time $\omega$ approaches infinity, we obtain the lower bound of the limiting probability $\pi_R$ from the characteristic equation (\ref{Eq14}) as follows:
\begin{equation}\label{Eqlimit}
  \pi_R \rightarrow \frac{h}{1+h} ~~\text{as}~~ \omega \rightarrow \infty.
\end{equation}
\end{IEEEproof}

The above corollary implies that the limiting probability $\pi_R$ cannot be further reduced by increasing a sufficiently large maximum waiting time $\omega$. The simulation result displayed in Fig. \ref{fig7c} confirms the fact that the limiting probability $\pi_R$ converges to the following lower bound given by (\ref{Eq31}):
\[ \frac{h}{1+h} = \frac{\frac{1}{180}}{1+\frac{1}{180}} \approx 0.0055, \]
when the maximum waiting time is set to equal $\omega=320 \mu s$, which is only slightly larger than the minimal requirement $\omega_{-1}=318 \mu s$ of the stable region, regardless of the initial fraction of ONUs in state $R$.

The limiting probability $\pi_R$ versus attempt probability $h$ is plotted in the Fig. \ref{fig9_Pi_R_h}, which shows that $-\alpha=eh$ almost coincides with the upper bound $-\frac{W_0(\alpha)}{1-W_0(\alpha)}$ when the attempt probability $h$ is small, especially when it is less than $1/16$. Note that the limiting probability $\pi_R$ displayed in Fig. \ref{fig9_Pi_R_h} exhibits similar behavior of $\omega_{-1}$ depicted in Fig. \ref{fig8_omega_h}. In the region $h<1/16$, the system is strictly stable and the limiting probability $\pi_R$ approaches the lower bound. When $h>1/16$, by contrast, the limiting probability $\pi_R$ quickly approaches the upper bound.

\section{Registration Delay}\label{section4}
In this section, we analyze the registration delay experienced by an ONU attempting to register. The registration delay, denoted by $d$, is defined as the duration from the time an ONU is turned on until it is successfully registered. The components of a typical registration delay are illustrated in Fig. \ref{fig10}. An ONU switched on at time $t_{on}$ can only make its registration attempt at the beginning of the next discovery window, denoted by $t_{next}$. The delay between $t_{on}$ and $t_{next}$ is called \emph{residual waiting delay}, denoted by $d_r$. If the attempt fails, the ONU tries again in the next registration cycle. We define the \emph{attempting delay}, denoted by $d_a$, as the duration of the registration cycles elapsed before the successful registration is achieved. As shown in Fig. \ref{fig10}, the registration delay is the sum of these two components: $d=d_r+d_a$, and the average registration delay is given by:
\begin{equation}\label{Eq33}
  E[d]=E[d_r]+E[d_a].
\end{equation}
We first estimate the mean residual waiting time in the following lemma.
\begin{lemma}\label{lemma3}
  The approximate mean residual waiting time is given by:
\begin{equation}\label{Eq34}
  E[d_r] \cong \frac{T}{2}.
\end{equation}
\end{lemma}
\begin{IEEEproof}
   Recall that $T$ is the registration cycle time. We consider the time interval $Z=T-d_r$, which is the interval between the beginning of the last discovery window, denoted by $t_{last}$, and the time that the ONU is turned on, denoted by $t_{on}$, as illustrated in Fig. \ref{fig10}. Also, the unregistered ONU may be in state $F$ or in state $A$ at $t_{last}$, and it shifts to state $R$ before or at time $t_{next}$. Thus, the distribution of $Z$ can be derived as follows:
\begin{align}\label{Eq35}
  P\{Z < z\} =& P\{(Z < z), (F \rightarrow R)\}    \nonumber \\
              & +P\{(Z < z) , (A \rightarrow R)\}  \nonumber \\
             =& P\{Z < z|F \rightarrow R\} P\{F \rightarrow R\}   \nonumber \\
              & +P\{Z < z|A \rightarrow R\} P\{A \rightarrow R\},
\end{align}
where the event $i \rightarrow R$, for $i\in \{F,A\}$, indicates that the ONU is in state $i$ at time $t_{last}$ and it engages in the registration process at time $t_{next}$. In the steady-state of the system, we have:
\[ P\{F \rightarrow R\}=\frac{\pi_F P_{F,R}}{\pi_F P_{F,R}+\pi_A P_{A,R}},\]
and
\[ P\{A \rightarrow R\}=\frac{\pi_A P_{A,R}}{\pi_F P_{F,R}+\pi_A P_{A,R}}.\]
Since $\tau_A \gg T$ and $\tau_F \gg T$, from (\ref{Eq2}), (\ref{Eq4}), and the approximation A1 given in Lemma \ref{lemma2}, we have:
\[\frac{P_{A,R}}{P_{F,R}} = \frac{\bigg{(} 1-e^{-\frac{T}{\tau_A}}\bigg{)} p_{rer}}{1-e^{-\frac{T}{\tau_F}}} \approx 0. \]
That is, the transition probability $P_{A,R}$ is negligible when it is compared to $P_{F,R}$, which implies $P\{F \rightarrow R\} \approx 1$ and $P\{A \rightarrow R\} \approx 0$. We then immediately obtain the following approximation from (\ref{Eq35}):
\begin{equation}\label{Eq36}
  P\{Z < z\} \cong P\{Z <z |F \rightarrow R\}.
\end{equation}
Given that the transition is from state $F$ to state $R$, the duration $z$ is the conditional power-off holding time of the ONU. Thus, the conditional probability (\ref{Eq36}) can be expressed as follows:
\begin{equation}\label{Eq37}
  P\{Z<z|F \rightarrow R\}=\frac{\int_0^z \frac{1}{\tau_F}e^{-\frac{t}{\tau_F}dt}}{\int_0^T \frac{1}{\tau_F}e^{-\frac{t}{\tau_F}}dt}=\frac{1-e^{-\frac{z}{\tau_F}}}{1-e^{-\frac{T}{\tau_F}}} \cong \frac{z}{T}.
\end{equation}
The last approximation is because $\tau_F \gg T > z$. Furthermore, we know from (\ref{Eq37}) that the random variable $Z$ is approximately uniformly distributed in the interval $[0, T]$ with mean $E[Z] \cong T/2$. As a result, the average residual waiting time is given by:
\[E[d_r]=E[T-Z] \cong \frac{T}{2}.\]
\end{IEEEproof}

Let $k (k=1,2,3,\cdots)$ be the number of discovery windows elapsed until an attempting ONU could register successfully. When $k=1$, the attempting delay of the ONU is the duration of a single discovery window. When $k=2$, the attempting delay of the ONU consists of a registration cycle and a single discovery window. In general, as illustrated in Fig. \ref{fig10}, the ONU spends $k-1$ registration cycles and a single discovery window on the registration. Hence, the attempting delay is given by:
\begin{equation}\label{Eq38}
  d_a=M+(k-1)T,
\end{equation}
where $M$ is the discovery window size defined in Section \ref{subsection2A}. Summarizing the above discussions, the mean registration delay is provided in the following theorem.
\begin{theorem}\label{theorem2}
  Suppose the maximum waiting time is selected in the stable region $\omega > \omega_{-1} $ or the saturated region $\omega < \omega_0$. The average registration delay that an ONU experiences in the registration processes is given by:
\begin{equation}\label{Eq39_Delay}
  E[d] \cong \Big{[} \frac{\pi_R}{(1-\pi_R)h} - \frac{1}{2} \Big{]}T.
\end{equation}
If the registration protocol is stable, the average delay is upper bounded by
\begin{equation}\label{Eq40_delay_bound}
  E[d] \leq 6.89T.
\end{equation}
\end{theorem}
\begin{IEEEproof}
We know from (\ref{Eq38}) that the average attempting delay is given by:
\begin{equation}\label{Eq41}
  E[d_a] = M+(E[k]-1)T \cong (E[k]-1)T,
\end{equation}
where the discovery window size $M$ can be ignored because it is typically much smaller than $T$. Recall that, in the steady-state of EPON, an ONU attempting to register succeeds in a discovery window with probability $p_{suc}$. Thus, $k$ is a geometric random variable with parameter $p_{suc}$. From (\ref{Eq14}), we have:
\begin{equation}\label{Eq42}
  E[k]=\frac{1}{p_{suc}}=\frac{\pi_R}{(1-\pi_R)h}.
\end{equation}
It follows from (\ref{Eq41}) and (\ref{Eq42}) that:
\begin{equation}\label{Eq43}
  E[d_a] \cong \Big{[} \frac{\pi_R}{(1-\pi_R)h} -1 \Big{]}T.
\end{equation}
Substituting (\ref{Eq34}) and (\ref{Eq43}) into (\ref{Eq33}), we immediately obtain (\ref{Eq39_Delay}).

If the system is stable,  we know from Corollary \ref{corollary2} that the limiting probability $\pi_R$ is upper bounded by $-\frac{W_0(\alpha)}{1-W_0(\alpha)}$. Moreover, we know from (\ref{Eq39_Delay}) that the mean delay is upper bounded by:
\begin{align}\label{Eq44_bound}
  E[d] & \cong  \Big{[} \frac{\pi_R}{(1-\pi_R)h} - \frac{1}{2} \Big{]}T  \leq  \Bigg{\{} \frac{-\frac{W_0(\alpha)}{1-W_0(\alpha)}}{\Big{[} 1 + \frac{W_0(\alpha)}{1-W_0(\alpha)} \Big{]}h} - \frac{1}{2} \Bigg{\}}T \nonumber \\
       & =      \Big{[} -\frac{W_0(\alpha)}{h} - \frac{1}{2} \Big{]}T = \Big{[} \frac{eW_0(\alpha)}{-eh} - \frac{1}{2} \Big{]}T \nonumber \\
       & =      \Big{[} \frac{eW_0(\alpha)}{\alpha} - \frac{1}{2} \Big{]}T.
\end{align}
According to the definition of the Lambert $W$ function $\alpha = W_0 (\alpha) e^{W_0 (\alpha)}$ given in Appendix \ref{appendixW}, we have:
\begin{align}\label{Eq45}
  E[d] & \leq \Big{[} \frac{eW_0(\alpha)}{\alpha} - \frac{1}{2} \Big{]}T =    \Big{[} \frac{eW_0(\alpha)}{W_0 (\alpha) e^{W_0 (\alpha)}} - \frac{1}{2} \Big{]}T \nonumber \\
       & =    \Big{[} e^{1-W_0(\alpha)} - \frac{1}{2} \Big{]}T  \leq \Big{(} e^2 -\frac{1}{2} \Big{)} \cong 6.89T.
\end{align}
\end{IEEEproof}

From Corollary \ref{corollary2}, we know that the limiting probability $\pi_R$ is close to the lower bound if the system is strictly stable. Thus, for a large $\omega>\omega_{-1}$ and a sufficiently small $h$ in the strictly stable region, substituting (\ref{Eqlimit}) into (\ref{Eq39_Delay}), we have the following ideal mean delay:
\begin{equation}\label{EqIdeal}
  E[d] \cong \Big{[} \frac{\pi_R}{(1-\pi_R )h} - \frac{1}{2} \Big{]} T \cong \Bigg{[} \frac{\frac{h}{1+h}}{\big{(}1-\frac{h}{1+h}\big{)}h}- \frac{1}{2} \Bigg{]T}=\frac{T}{2}.
\end{equation}

\begin{figure}[t]
\centering
\includegraphics[width=0.45\textwidth]{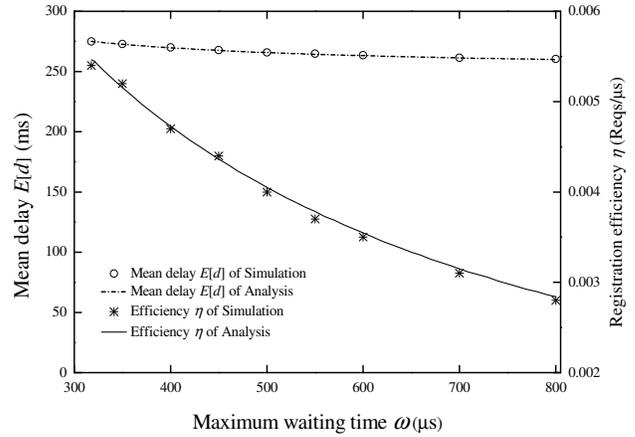}
\caption{$E[d]$ and $\eta$ versus $\omega$ when $\omega \geq \omega_{-1}$.}\label{fig11}
\end{figure}

On the other hand, the registration delay is very large if the maximum waiting time $\omega$ is in the saturated region $\omega < \omega_0$. When $\omega$ is very small and the limiting probability $\pi_R$ approaches 1, the average registration delay $E[d]$  given by (\ref{Eq39_Delay}) may not be bounded. This property of an unstable registration protocol has been confirmed by our simulation results, which demonstrated the fact that an ONU cannot successfully register even after tens of thousands of registration cycles when $\omega < \omega_0$.

Intuitively, the average registration delay $E[d]$ decreases with the increase of the maximum waiting time $\omega$ in the stable region $\omega > \omega_{-1}$. This property is confirmed by the simulation results and the theoretical results plotted in Fig. \ref{fig11}, where the parameters are the same as those used in Fig. \ref{fig6}. The predominate delay component is $E[k]$, the average number of discovery windows elapsed until the successful registration is achieved. With the increasing of $\omega$, the probability $p_{suc}$ that a registration request is successful in a discovery window also increases, and thus $E[k]$ decreases. However, Fig. \ref{fig11} also shows that the reduction of the delay by increasing $\omega$ is marginal. For instance, when $\omega$ increases from $317.8 \mu s$ to $800 \mu s$, more than doubling, the average delay $E[d]$ only decreases by 5.5\%, from $275 ms$ to $260 ms$. This can be explained as follows. In the strictly stable region, the limiting probability $\pi_R$ is already very small and, thus, can only be reduced slightly by increasing $\omega$, as demonstrated by (\ref{Eqlimit}). It follows from (\ref{Eq42}) that the improvement in $p_{suc}$, or equivalently the reduction of $E[k]$, is insignificant. Consequently, the reduction of $E[d]$ by increasing $\omega$ is negligible, as predicted by (\ref{EqIdeal}).

In contrast, the increase of $\omega$ not only shrinks the bandwidth available for normal transmission, but also affects the registration efficiency. In \cite{1kramer2005,9cui2012throughput}, this indicator, denoted by $\eta$, is defined as the ratio of the number of successful registrations to the discovery window size and is given as follows:
\begin{align}\label{EqEff}
  \eta =& \frac{N\pi_R p_{suc}}{M} = \frac{Nh (1-\pi_R)}{2Q+D} \cong  \frac{Nh}{2Q+\omega +L},
\end{align}
where the second equality is obtained from (\ref{Eq14}) and the last approximation is due to the condition $\pi_R \ll 1$. A comparison of $E[d]$ in (\ref{EqIdeal}) and $\eta$ in (\ref{EqEff}) shows that the efficiency $\eta$ drops much faster than then mean delay $E[d]$ in respect to the increase of $\omega$. This point is also confirmed by the simulation results displayed in Fig. \ref{fig11}, which shows that the efficiency $\eta$ is reduced by almost half when $\omega$ increases from $317.8 \mu s$ to $800 \mu s$. Thus, a marginal reduction of the delay is achieved at the expense of a substantial degradation of registration efficiency $\eta$.

\section{Conclusion}\label{conclusion}

In this paper, we analyze the stability and delay of the EPON registration protocol. We first establish a model of the subscribers by using a discrete-time Markov chain, and then derive the characteristic function to delineate the throughput of the registration process. Solving the characteristic function, we obtain the region of the maximum waiting time to make the registration protocol stable. If the maximum waiting time is selected from this region, we show that a stable registration throughput and a bounded registration delay can be guaranteed. Our results also indicate that it is unnecessary to arbitrarily enlarge the maximum waiting time as long as it is in the stable region since the improvement of registration delay is marginal, yet the reduction of registration efficiency is quite significant.

\appendices
\section{Lambert $W$ function.}\label{appendixW}
\begin{figure}[t]
\centering
\includegraphics[width=0.40\textwidth]{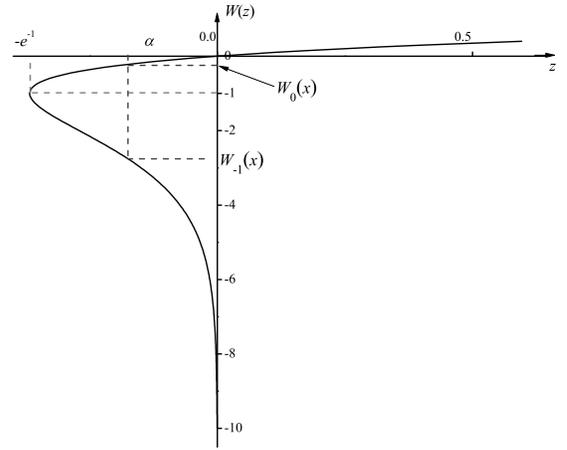}
\caption{The Lambert $W$ function.}\label{figW}
\end{figure}

The Lambert $W$ function $W(z)$, which was first considered by J. Lambert \cite{14corless1996lambertw}, is defined as the function satisfying:
\begin{equation}\label{EqB1}
  W(z)e^{W(z)}=z.
\end{equation}
The Lambert $W$ function $W(z)$ is plotted in Fig. \ref{figW}. If $z$ is real and $-e^{-1}<z<0$, $W(z)$ has two real values: the principal branch $W_0(z)\in [-1,0)$, and the other branch $W_{-1}(z) \in (-\infty,-1]$. If $z$ is real and $z>0$, $W(z)$ has only a branch $W_0(z) \in [0,\infty)$.

Ref. \cite{14corless1996lambertw} shows that, according to the Lagrange inversion theorem, $W_0(z)$ has the following series expansion:
\begin{equation*}
  W_0(z)=\overset{\infty}{\underset{i=1}\sum}{\frac{(-i)^{i-1}}{i!}z^i}=z-z^2+\frac{3}{2}z^3-\frac{8}{3}z^4+\cdots,
\end{equation*}
and $W_{-1}(z)$ has the following series expansion:
\begin{equation*}
  W_{-1}(z)=\overset{\infty}{\underset{i=0}\sum}{\mu_i y^i}=-1+y-\frac{1}{3} y^2+\frac{11}{72} y^3-\frac{43}{540} y^4+\cdots,
\end{equation*}
where $y=-\sqrt{(2(ez+1)}$. The coefficient $\mu_i$ is given in \cite{14corless1996lambertw}.

\section{Solutions of the characteristic equation}\label{appendixchar}
In this appendix, we explore the solutions of the following characteristic equation that Theorem \ref{theorem1} establishes:
\begin{equation}\label{EqA1}
  (1-\pi_R)h=\pi_R e^{-\frac{2LN}{\omega}\pi_R},
\end{equation}
Define the auxiliary function:
\begin{equation}\label{EqA2}
  F(x,\omega)=(1-x)h-xe^{-\frac{2LN}{\omega}x},
\end{equation}
Furthermore, this auxiliary function is denoted as $F(x,\omega)=f_\omega(x)$ for a fixed $\omega$, and $F(x,\omega)=g_x(\omega)$ for a fixed $x$. It is easy to show that the function $f_\omega(x)$ is continuous at all real $x$, and we have:
\begin{equation}\label{EqA3}
    \begin{cases}
        f_\omega(x)>0, & \text{when} ~x \leq 0 \\
        f_\omega(x)<0, & \text{when} ~x \geq 1
   \end{cases},
\end{equation}
which implies that $f_\omega (x)=0$ has at least one real root, and all real roots of this equation lie in the interval $[0,1]$. From the definition of (\ref{EqA2}), all real roots of $f_\omega (x)=0$ are valid solutions for the characteristic equation (\ref{EqA1}) of $\pi_R$. In this appendix, we first enumerate the number of roots of $f_\omega (x)=0$ for a given maximum waiting time $\omega$, then we prove that the only real root of $f_\omega (x)=0$ must lie in the interval $\big{[}0,-\frac{W_0(\alpha)}{1-W_0(\alpha)}\big{]}$  when $\omega > \omega_{-1}$.

\begin{table*}[t]
  \centering
  \caption{The number of roots of the equation $f_\omega(x)=0$}\label{table1}
  \begin{tabular}{|c|c|c|c|c|}
    \hline
    \multirow{2}{*}{Extreme Values of $f_\omega(x)$} & Maximum waiting time & Monotonically decreasing & Monotonically increasing & Monotonically decreasing \\
                                         &          $\omega$                & interval $[0,x_0]$ & interval $[x_0,x_1]$ & interval $[x_1,1]$ \\
    \hline
    $0 < f_\omega(x_0) < f_\omega (x_1)$ & $\omega < \omega_0 < \omega_{-1}$ &          -         &          -           & \text{one root}    \\
    \hline
    $0 = f_\omega(x_0) < f_\omega (x_1)$ & $\omega = \omega_0 < \omega_{-1}$ &        $x_0$       &        $x_0$         & \text{one root}    \\
    \hline
    $f_\omega(x_0) < 0 < f_\omega (x_1)$ & $\omega_0 < \omega < \omega_{-1}$ &   \text{one root}  &    \text{one root}   & \text{one root}    \\
    \hline
    $f_\omega(x_0) < f_\omega (x_1) = 0$ & $\omega_0 < \omega = \omega_{-1}$ &   \text{one root}  &        $x_1$         &       $x_1$        \\
    \hline
    $f_\omega(x_0) < f_\omega (x_1) < 0$ & $\omega_0 < \omega_{-1} < \omega$ &   \text{one root}  &          -           &        -          \\
    \hline
  \end{tabular}
\end{table*}

\subsection{Determine the number of roots of $f_\omega(x)=0$}
For a fixed $\omega$, taking the derivative of (\ref{EqA2}) with respect to $x$, we have:
\begin{equation}\label{EqA4}
  f'_\omega(x)= \frac{\partial F(x,\omega)}{\partial x}=-h- \bigg{(} 1-\frac{2LNx}{\omega}\bigg{)}e^{-\frac{2LN}{\omega}x}.
\end{equation}
Solving $f'_\omega (x)=0$, we obtain the following two roots:
\begin{equation}\label{EqA5}
  x_0=\frac{\omega [1-W_0(\alpha)]}{2LN},
\end{equation}
and
\begin{equation}\label{EqA6}
  x_1=\frac{\omega [1-W_{-1}(\alpha)]}{2LN},
\end{equation}
where $\alpha=-eh$ is a constant, and $W_0(\alpha)$ and $W_{-1}(\alpha)$ are the two possible values of the Lambert $W$ functions [14] as described in Appendix \ref{appendixW}. Substituting (\ref{EqA5}) and (\ref{EqA6}) into (\ref{EqA2}), we obtain the following two possible extreme values of $f_\omega(x)$:
\begin{equation}\label{EqA7}
  f_\omega(x_0)= h+\frac{\omega h [1-W_0(\alpha)]^2}{2LNW_0(\alpha)}=h(1-\frac{\omega}{\omega_0}),
\end{equation}
and
\begin{equation}\label{EqA8}
  f_\omega(x_1)= h+\frac{\omega h [1-W_{-1}(\alpha)]^2}{2LNW_{-1}(\alpha)}=h(1-\frac{\omega}{\omega_{-1}}),
\end{equation}
where $\omega_0$  and $\omega_{-1}$ are parameters defined by (\ref{Eq23}) and (\ref{Eq24}), respectively.

\begin{figure}[t]
\centering
    \subfigure[]{\label{figA1a}
    \begin{minipage}[c]{0.32\textwidth}
    \centering
    \includegraphics[width=1\textwidth]{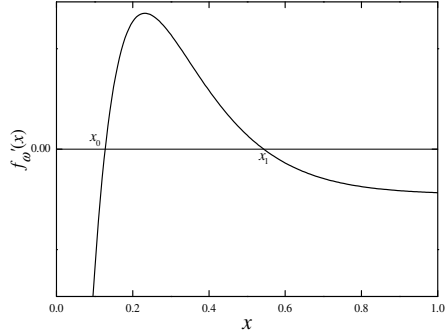}
    \vspace{0.001\textwidth}
    \end{minipage}%
    }%注意这个”%”绝对不能省，可以试试不打%的效果

    \subfigure[]{\label{figA1b}
    \begin{minipage}[c]{0.32\textwidth}
    \centering
    \includegraphics[width=1\textwidth]{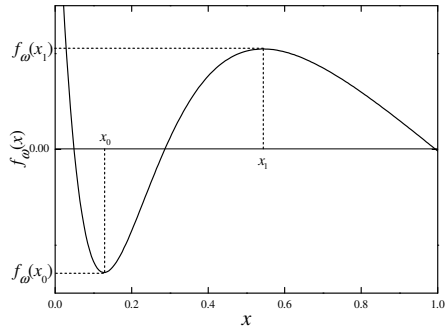}
    \vspace{0.001\textwidth}
    \end{minipage}%
    }%注意这个”%”绝对不能省，可以试试不打%的效果
\caption{(a) The extreme value of $f'_\omega(x)$; (b) The two extreme values of $f_\omega(x)$.}\label{figA1}
\end{figure}

Since we know from (\ref{EqA3}) that all real roots of $f_\omega(x)=0$ lie in the interval $[0,1]$, the function $f_\omega (x)$ exhibits the following properties in the three sub-intervals $[0,x_0]$, $[x_0,x_1]$, and $[x_1,1]$:
\begin{enumerate}
  \item [1)] In $[0,x_0]$,~$f'_\omega (x)<0$ and $f_\omega (x)$ is monotonically decreasing;
  \item [2)] In $[x_0,x_1]$,~$f'_\omega (x)>0$ and $f_\omega (x)$ is monotonically increasing;
  \item [3)] In $[x_1,1]$,~$f'_\omega (x)<0$ and $f_\omega (x)$ is monotonically decreasing again.
\end{enumerate}
It follows that (\ref{EqA7}) and (\ref{EqA8}) are the local minima and local maxima, respectively, of the function $f_\omega (x)$, as illustrated in Fig. \ref{figA1b}.

\begin{figure}[t]
\centering
\includegraphics[width=0.45\textwidth]{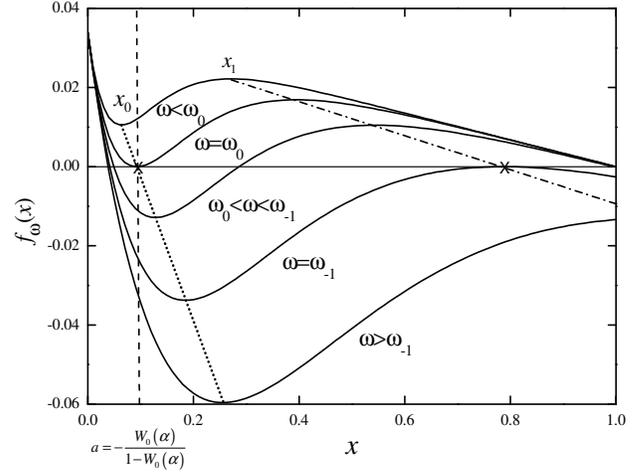}
\caption{The function $f_\omega(x)$ for different values of $\omega$.}\label{figA2}
\end{figure}

The number of roots of the function $f_\omega (x)$ is determined by the two extreme values $f_\omega (x_0)$ and $f_\omega (x_1)$ given by (\ref{EqA7}) and (\ref{EqA8}), respectively. Since $f_\omega (x)$ is monotonically increasing in the interval $[x_0,x_1]$, we know that $f_\omega (x_0) < f_\omega (x_1)$, or equivalently  $\omega_0<\omega_{-1}$, always holds. For different values of the maximum waiting time $\omega$, as illustrated in Fig. \ref{figA2}, the number of solutions of the characteristic equation (\ref{EqA1}) for different values of $\omega$ is summarized in Table \ref{table1}.

\subsection{Proof of Corollary \ref{corollary2}}
Suppose that $\omega_0 < \omega_{-1} <\omega$. We know that the function $f_\omega (x)=0$ has a unique real root in the interval $[0,x_0 ]$, where $f'_\omega (x_0 )=0$ and $f_\omega (x)$ has a local minima at $x_0$. We want to show that this root of $f_\omega (x)=0$ lies in the interval $\big{[}0,-\frac{W_0(\alpha)}{1-W_0(\alpha)}\big{]}$.

Recall that $f_\omega (0)>0$ for all $\omega$, and $f_\omega (x)$ has a local minima at:
\[ x_0 = \frac{\omega [1-W_0(\alpha)]}{2LN}. \]
When $\omega=\omega_0$, denote this local minima as $x_0=a$ and $f_{\omega_0} (a)=0$. From (\ref{Eq23}), we have:
\[ a=\frac{ \omega_0 [1-W_0 (\alpha)]}{2LN} =-\frac{W_0(\alpha)}{1-W_0(\alpha)} > 0. \]
For a fixed $x$, taking the derivative of (\ref{EqA2}) with respect to $\omega$, we obtain:
\begin{equation*}
  g'_x (\omega) = \frac{\partial F (x, \omega)}{\partial \omega} = -\frac{2LNx^2}{\omega ^2}e^{-\frac{2LN}{\omega}x} <0.
\end{equation*}
Thus, the function $g_x (\omega)$ is monotonically decreasing with respect to $\omega$, for any fixed $x$. In particular, when $\omega_0 < \omega_{-1} < \omega$, we have:
\begin{equation*}
  g_{x=a} (\omega)< g_{x=a} (\omega_0) = F (a, \omega_0) = f_{\omega_0} (a) =0,
\end{equation*}
which, however, implies:
\begin{equation}\label{EqA9}
  f_{\omega} (a) = F (a, \omega) = g_{x=a} (\omega) < 0.
\end{equation}
Since we know $f_\omega (0)>0$, it follows from (\ref{EqA9}) that the unique real root of $f_\omega (x)=0$ must lie in the interval $[0,a]= [ 0,-\frac{W_0 (\alpha)}{(1-W_0(\alpha)} ]$, as illustrated in Fig. \ref{figA2}.

\bibliography{IEEEabrv,Bib_EPON_stability}
\end{document}